\documentclass[12pt]{article}
\usepackage{latexsym,amsmath,amsfonts,amssymb}
\newtheorem{Theorem}{Theorem}
\newtheorem{Def}{Definition}
\def\p{\partial}
\def\nn{\nonumber}
\author{Renat Zhdanov\thanks{E-mail:\ renat.zhdanov@bio-key.com}\\
BIO-key International, Eagan, MN, USA}
\title{Nonlocal symmetries of systems of evolution equations.}
\date{}

\begin{document}
\maketitle

\begin{abstract}
We prove that any potential symmetry of a system of evolution
equations reduces to a Lie symmetry through a nonlocal
transformation of variables. Based on this fact is our method of
group classification of potential symmetries of systems of evolution
equations having non-trivial Lie symmetry. Next, we modify the above
method to generate more general nonlocal symmetries, which yields a
purely algebraic approach to classifying nonlocal symmetries of
evolution type systems. Several examples are considered.
\end{abstract}
\section{Introduction}

Lie symmetries and their various generalizations have become an
inseparable part of the modern physical description of wide range of
phenomena of nature from quantum physics to hydrodynamics. Such
success of a purely mathematical theory of continuous groups
developed by Sophus Lie in 19th century \cite{lie1} is explained by
the remarkable fact that the overwhelming majority of mathematical
models of physical, chemical and biological processes possess
nontrivial Lie symmetry.

One can even argue that this very property, invariance under Lie
symmetries, distinguish the popular models of mathematical and
theoretical physics from a continuum of possible models in the form
of differential or integral equations (see, e.g., \cite{fus1,zhd1}).
Based on this observation is the symmetry selection principle
stating that if an equation describing some physical process
contains arbitrary elements, then the latter should be so chosen
that the resulting model possesses the highest possible symmetry. In
this sense Lie theory effectively predicts which equation is the
best candidate to serve as a mathematical model of a specific
physical, chemical or biological process.

The process of choosing from a prescribed class of differential
equations those enjoying the highest Lie symmetry is called group
classification. In the case when non-Lie symmetries are involved,
the more general term, symmetry classification, is used.

In this paper we study symmetries of systems of evolution equations
in one spatial variable
\begin{equation}
\label{0.1}
{\mathbf u}_t={\mathbf f}(t,x,{\mathbf u},{\mathbf
u}_1,\ldots,{\mathbf u}_n),
\end{equation}
where ${\mathbf u}=\{u^1(t,x),u_2(t,x),\ldots,u^m(t,x)\}$,\
${\mathbf u}_{i+1}=\p{\mathbf u}_i/\p x$, $n\ge 2$, $m\ge 2$.

There is a lot papers devoted to group classification of different
subclasses of the class of partial differential equations (PDEs) of
the form (\ref{0.1}) (see, e.g. \cite{ibr1}-\cite{nik3} and the
references therein). The major tool utilized in these studies is the
infinitesimal Lie approach enabling to reduce the problem of
exhaustive description of Lie transformation groups admitted by
(\ref{0.1}) to integrating some linear system of PDEs (for further
details, see \cite{ovs1}-\cite{blu1}).

However, with all importance and power of the traditional Lie
approach, it does not provide all the answers to the mounting
challenges of the modern nonlinear physics. By this very reason
there were numerous attempts of generalize Lie symmetries so that
the generalized symmetries retain the most important features of Lie
symmetries and allow for broader scope of applicability. The natural
move in this direction would be to allow for the coefficients of
infinitesimal generators of Lie symmetries to contain not only
independent and dependent variables and their derivatives but
integrals of dependent variables, as well. In this way, the so
called nonlocal symmetries were introduced into mathematical
physics.

The concept of nonlocal symmetry of linear PDEs is well understood
by now (see, e.g., \cite{fus2}). However, this is not the case for
nonlinear equations. The problem of developing regular methods for
constructing nonlocal symmetries of nonlinear PDEs is still waiting
for its Sophus Lie. On the other hand, there is a number of results
on nonlocal symmetries for specific equations. One of the possible
approaches to construction of nonlocal symmetries has been suggested
by Bluman \cite{blu2}-\cite{blu4}. He put forward the concept of
potential symmetry, which is a special case of nonlocal symmetry.
The basic idea of the method for constructing potential symmetries
of PDEs can be formulated in the following way. Consider evolution
equation
\begin{equation}
\label{0.2}
u_t=f(t,x,u,u_1,\ldots,u_n).
\end{equation}
Suppose that it can be rewritten in the form of a conservation law
\begin{equation}
\label{0.3}
\frac{\p}{\p t}\Bigl(G(t,x,u)\Bigr) = \frac{\p}{\p
x}\Bigl(F(t,x,u,u_1,\ldots,u_{n-1})\Bigr).
\end{equation}
By force of (\ref{0.3}) we can introduce new dependent variable
$v=v(t,x)$ and rewrite equation (\ref{0.1}) as follows
\begin{equation}
\label{0.4}
v_x=G(t,x,u),\quad v_t=F(t,x,u,u_1,\ldots,u_{n-1}).
\end{equation}
Now if system of two equations (\ref{0.4}) admits Lie symmetry such
that at least one of the coefficients of its infinitesimal operators
depends on $v = \p_x^{-1}G(t,x,u)$, then this symmetry is the
nonlocal symmetry for the initial evolution equation (\ref{0.2}).
Here $\p_x^{-1}$ is the inverse of $\p_x$, i.e., $\p_x\,\p_x^{-1}$
$\equiv$ $\p_x^{-1}\,\p_x$ $\equiv 1$. This nonlocal symmetry is
also called potential symmetry of (\ref{0.2}).

Pucci and Saccomandi proved that potential symmetries can be derived
using non-classical symmetries of the (\ref{0.2}). Recently, we
established much stronger assertion by associating potential
symmetries with classical (contact) symmetries \cite{zhd2}. More
precisely, we proved that any potential symmetry of evolution
equation (\ref{0.2}) can be reduced to contact symmetry by a
suitable nonlocal transformation of dependent and independent
variables. As a consequence, one can obtain exhaustive description
of potential symmetries of (\ref{0.2}) through classification of
contact symmetries of PDEs of the form (\ref{0.2}).

Some applications of potential symmetries to specific subclasses of
(\ref{0.2}) can be found in \cite{sac3}-\cite{tor1}.

In the present paper we generalize results of \cite{zhd2} for system
of evolution equations (\ref{0.1}) and prove that any potential
symmetry of the system in question reduces to classical Lie symmetry
under a suitable nonlocal transformation of dependent and
independent variables (Sections 1, 2). Next, we suggest in Section 3
a more general approach to constructing nonlocal symmetries that
goes far beyond of the concept of potential symmetries. This
approach enables generating systems of evolution equations
associated with a given system of the system (\ref{0.1}), provided
the latter admits non-trivial Lie symmetry. We give several examples
of application of the approach in Section 3.

\section{Conservation law representation and classical symmetries}

\begin{Def}
We say that system (\ref{0.1}) admits complete conservation law
representation (CLR) if it can be written in the form
\begin{equation}
\label{1.1}
\frac{\p}{\p t}\Bigl({\mathbf G}(t,x,{\mathbf u})\Bigr)
= \frac{\p}{\p x} \Bigl({\mathbf F}(t,x,{\mathbf u},{\mathbf
u}_1,\ldots,{\mathbf u}_{n-1})\Bigr).
\end{equation}
Here ${\mathbf u},{\mathbf F},{\mathbf G}$ are $m$-component
vectors.
\end{Def}
\begin{Def}
We say that system (\ref{0.1}) admits partial CLR if it can be
written in the form
\begin{eqnarray}
&&\frac{\p}{\p t}\Bigl({\mathbf F}(t,x,{\mathbf u},{\mathbf
w})\Bigr) = \frac{\p}{\p x}\Bigl({\mathbf G}(t,x, {\mathbf
u},{\mathbf u}_1,\ldots,{\mathbf u}_{n-1}, {\mathbf w},{\mathbf
w}_1,\ldots,{\mathbf w}_{n-1})\Bigr),\nn\\
&&{\mathbf w}_t = {\mathbf H}(t,x,{\mathbf u},{\mathbf
u}_1,\ldots,{\mathbf u}_{n}, {\mathbf w},{\mathbf
w}_1,\ldots,{\mathbf w}_{n}).\label{1.2}
\end{eqnarray}
Here ${\mathbf u},{\mathbf F},{\mathbf G}$ and ${\mathbf w},{\mathbf
H}$ are $r$-component and $m-r$-component vectors, respectively.
\end{Def}

Below we present theorems that provide exhaustive characterization
of conservation law representability in terms of classical Lie
symmetries. We give the detailed proof of the assertion regarding
complete CLR, the case of partial CLR is handled in a similar way.

\begin{Theorem}
System (\ref{0.1}) admits complete CLR if and only if it is
invariant under $m$-dimensional commutative Lie algebra
$\mathcal{L}_m=\langle e_1,$ $\ldots,$ $e_m\rangle$, where
\begin{equation}
\label{1.3}
e_i=\xi_i(t,x,{\mathbf u})\p_x + \sum_{j=1}^m\,\eta_i^j(t,x,{\mathbf
u})\p_{u^j},
\end{equation}
and besides,
\begin{equation}
\label{1.4}
 {\rm rank}\,\left(
\begin{array}{llll}
\xi_1 & \eta_1^1 & \ldots & \eta_1^m\\[2mm]
\vdots& \vdots   & \vdots & \vdots  \\[2mm]
\xi_m & \eta_m^1 & \ldots & \eta_m^m
\end{array}
\right) = m.
\end{equation}
\end{Theorem}
{\bf Proof.} Suppose system (\ref{0.1}) admits CLR (\ref{1.1}).
Introducing new $m$-component function
\begin{equation}
\label{1.5}
{\mathbf v}_x = {\mathbf G}(t,x,{\mathbf u})
\end{equation}
and eliminating ${\mathbf u}$ from (\ref{1.1}) we get
\begin{equation}
\label{1.6} {\mathbf v}_{xt}=\frac{\p}{\p x}\Bigl(\tilde{\mathbf
f}(t,x,{\mathbf v}_1,\ldots,{\mathbf v}_{n})\Bigr).
\end{equation}
Integrating the obtained system of PDEs with respect to $x$ yields
\begin{equation}
\label{1.7}
{\mathbf v}_{t}=\tilde{\mathbf f}(t,x,{\mathbf v}_1,\ldots,{\mathbf
v}_{n}).
\end{equation}
Note that the integration constant ${\mathbf w}(t)$ is absorbed into
the function ${\mathbf v}$. Evidently, system (\ref{1.7}) is
invariant under the commutative $m$-dimensional Lie algebra
${\mathcal L}_m$ $=\langle \p_{v^1},$ $\ldots,$ $\p_{v^m}\rangle$.
What is more, the coefficients of the basis elements of the algebra
${\mathcal L}_m$ satisfy condition (\ref{1.4}).

Let us prove now the that the inverse assertion is also true.
Suppose that (\ref{0.1}) admits Lie algebra ${\mathcal L}_m$
$=\langle e_1,$ $\ldots,$ $e_m\rangle$, whose basis elements have
the form (\ref{1.3}) and satisfy (\ref{1.4}). Then there is a change
of variables (see, e.g. \cite{ovs1})
$$
\bar t=t,\quad \bar x=X(t,x,{\mathbf u}),\quad \bar {\mathbf u} =
{\mathbf U}(t,x,{\mathbf u})
$$
reducing basis elements of ${\mathcal L}_m$ to the form
$e_i=\p_{{\bar u}^i},\ i=1,\ldots,m$. In what follows we drop the
bars.

Now (\ref{0.1}) necessarily takes the form
\begin{equation}
\label{1.8}
{\mathbf u}_t=\tilde{\mathbf f}(t,x,{\mathbf u}_1,\ldots,{\mathbf
u}_n).
\end{equation}
Differentiating (\ref{1.8}) with respect to $x$ and making the
(nonlocal) change of dependent variables ${\mathbf v}_x = {\mathbf
u} $, we finally get
$$
{\mathbf v}_t=\frac{\p}{\p x}\tilde{\mathbf f}(t,x,{\mathbf
v},{\mathbf v}_1,\ldots,{\mathbf v}_{n-1}),
$$
which completes the proof.\quad$\square$

\noindent {\bf Note 1.}\ The fact that symmetry operators, $e_1$,
$\ldots$, $e_m$ are of specific form (\ref{1.3}) is crucial for the
whole procedure of reducing a system of evolution equations to a
'conserved' form (\ref{1.1}). If a symmetry group generated by some
operator $e_i$ does not preserve the temporal variable, $t$ (which
means that the coefficient of $\p_t$ in $e_i$ is non-zero for some
$i$), then this operator cannot be reduced to the canonical form
$\p_{v^i}$ and the reduction routine does not work.

\begin{Theorem}
System (\ref{0.1}) admits partial CLR if and only if it is invariant
under $r$-dimensional commutative Lie algebra $\mathcal{L}_r=\langle
e_1,$ $\ldots,$ $e_r\rangle$, where
\begin{eqnarray}
&&e_i=\xi_i(t,x,{\mathbf u},{\mathbf w})\p_x +
\sum_{j=1}^m\,\eta_i^j(t,x,{\mathbf u},{\mathbf w})\p_{u^j}\nn\\
&&\quad + \sum_{j=1}^m\,\zeta_i^j(t,x,{\mathbf u},{\mathbf
w})\p_{w^j} \quad i=1,\ldots,r,\label{1.9}
\end{eqnarray}
and besides,
\begin{equation}
\nn
{\rm rank}\left(
\begin{array}{lllllll}
\xi_1 & \eta_1^1 & \ldots & \eta_1^m & \zeta_1^1 & \ldots & \zeta_1^m\\[2mm]
\vdots& \vdots   & \vdots & \vdots   & \vdots    & \vdots & \vdots \\[2mm]
\xi_r & \eta_r^1 & \ldots & \eta_r^m & \zeta_r^1 & \ldots & \zeta_r^m
\end{array}
\right) = r.
\end{equation}
\end{Theorem}

\section{Potential symmetries}

Potential symmetries of system of evolution equations (\ref{0.1})
appear in the same way as they do for a single evolution equation.
For simplicity, we consider the case of complete CLR. By force of
(\ref{1.1}) we can introduce the new dependent variable ${\mathbf
v}$, so that
\begin{equation}
\label{2.1}
{\mathbf v}_t = {\mathbf F}(t,x,{\mathbf u},{\mathbf
u}_1,\ldots,{\mathbf u}_{n-1}),\quad {\mathbf v}_x = {\mathbf
G}(t,x,{\mathbf u}).
\end{equation}
Note that ${\mathbf v}$ is nonlocal variable since ${\mathbf v}$ $=$
$\p_x^{-1}{\mathbf G}(t,x,{\mathbf u})$.

Suppose now that system (\ref{2.1}) admits Lie symmetry
\begin{eqnarray}
&&t' = T(t,x,{\mathbf u},{\mathbf v},\theta),\quad x' =
X(t,x,{\mathbf u},{\mathbf v},\theta),\nn\\ &&{\mathbf u}' =
{\mathbf U}(t,x,{\mathbf u},{\mathbf v},\theta),\quad {\mathbf v}' =
{\mathbf V}(t,x,{\mathbf u},{\mathbf v},\theta),\label{2.2}
\end{eqnarray}
such that one of the derivatives
\begin{equation}
\nn \frac{\p T}{\p v^i},\quad \frac{\p T}{\p v^i},\quad
\frac{\p\mathbf U}{\p v^i},\quad \frac{\p\mathbf V}{\p v^i},\quad
i=1,\ldots,m
\end{equation}
does not vanish identically. Rewriting group (\ref{2.2}) in terms of
variables $t,x,{\mathbf u}$ and taking into account that ${\mathbf
v} = {\partial}_x^{-1}{\mathbf u}$ yield the nonlocal symmetry of
the initial system of evolution equations (\ref{0.1}). This means,
in particular, that symmetry in question cannot be obtained within
the Lie infinitesimal approach. What we are going to prove, is that
this symmetry can be derived by regular Lie approach if the later is
combined with the nonlocal transformation of the dependent
variables.

Indeed, let system (\ref{0.1}) admit complete CLR (\ref{1.1}). In
addition, we suppose that (\ref{0.1}) possesses potential symmetry.
Making the nonlocal change of dependant variables, ${\mathbf u}$
$\rightarrow$ ${\mathbf v}$,
\begin{equation}
\label{2.3}
{\mathbf v}_x = {\mathbf G}(t,x,{\mathbf u}),\quad {\mathbf u} =
\widetilde{\mathbf G}(t,x,{\mathbf v}_x),\quad G(t,x,
\widetilde{\mathbf G}(t,x,{\mathbf v}_x))\equiv {\mathbf v}_x
\end{equation}
we rewrite (\ref{1.1}) in the form (\ref{1.6}). As initial system
(\ref{0.1}) admits a potential symmetry, system (\ref{2.1}) is
invariant under the Lie transformation group of the form
(\ref{2.2}).

Integrating (\ref{1.6}) with respect to $x$ yields system of
evolution equations
\begin{equation}
\label{2.4}
{\mathbf v}_{t}=\tilde{\mathbf f}(t,x,{\mathbf v}_1,\ldots,{\mathbf
v}_{n}).
\end{equation}
Next we rewrite Lie symmetry (\ref{2.2}) by eliminating ${\mathbf
u}$ according to (\ref{2.3}) which yields
\begin{eqnarray}
&&t' = T\Bigl(t,x,\widetilde{\mathbf G}(t,x,{\mathbf v}_x),{\mathbf
v},\theta\Bigr),\quad x' = X\Bigl(t,x,\widetilde{\mathbf
G}(t,x,{\mathbf v}_x),{\mathbf v},\theta\Bigr),\nn\\ &&{\mathbf v}'
= {\mathbf V}\Bigl(t,x,\widetilde{\mathbf G}(t,x,{\mathbf
v}_x),{\mathbf v},\theta\Bigr).\label{2.5}
\end{eqnarray}
By construction, Lie transformation group (\ref{2.5}) maps the set
of solutions of (\ref{2.4}) into itself. Consequently, (\ref{2.5})
is the Lie group of contact symmetries of system of evolution
equations (\ref{2.4}).

It is a common knowledge that any contact symmetry of a system of
PDEs boils down to the first prolongation of a classical symmetry
\cite{ibr2}. Consequently, the derivatives of $T, X, {\mathbf V}$
with respect to the third argument vanish identically and we get
\begin{equation}
\label{2.6}
t' = T(t,x,{\mathbf v},\theta),\quad x' = X(t,x,{\mathbf
v},\theta),\quad {\mathbf v}' = {\mathbf V}(t,x,{\mathbf v},\theta).
\end{equation}
This group is nothing else than the standard Lie symmetry group of
system (\ref{2.4}).

The same assertion holds true for the case of partial CLR.
\begin{Theorem}
Let system of evolution equations (\ref{0.1}) admit complete or
partial CLR and be invariant under a potential symmetry. Then there
exist a (nonlocal) change of variables mapping (\ref{0.1}) into
another system of the form (\ref{0.1}) so that the potential
symmetry of (\ref{0.1}) becomes the standard Lie symmetry of the
transformed system.
\end{Theorem}

This assertion is in fact a no-go theorem for potential symmetries
of systems of evolution equations. It states that the concept of
potential symmetry does not produce essentially new symmetries. The
system admitting potential symmetry is equivalent to the one
admitting standard Lie symmetry, which is the image of the potential
symmetry in question.

However, there is more to it. Theorem 3 imply the regular algorithm
for group classification system of nonlinear evolution equations
admitting nonlocal symmetries. Again, for the sake of simplicity, we
consider the case of complete CLR.

Indeed, let system of evolution equations (\ref{0.1}) be invariant
under $(m+1)$-dimensional Lie algebra $\mathcal{L}_{m+1}=\langle
e_1,$ $\ldots,$ $e_{m+1}\rangle$. Here $e_1$, $\ldots$, $e_m$ are
commuting operators of the form (\ref{1.3}) and their coefficients
satisfy constraint (\ref{1.4}). Basis operator $e_{m+1}$ is of the
generic form
$$
e_{m+1}=\tau(t,x,{\mathbf u})\p_t + \xi_i(t,x,{\mathbf u})\p_x +
\sum_{j=1}^m\,\eta_i^j(t,x,{\mathbf u})\p_{u^j}.
$$

Making an appropriate change of variables we can reduce the
operators $e_1$, $\ldots$, $e_m$ to the canonical forms, namely,
$e_i=\p_{u^i}$,\ $i=1,\ldots,m$. Then system (\ref{0.1}) necessarily
takes the form (\ref{2.4}).

Let (\ref{2.6}) be Lie transformation group generated by the
symmetry operator $e_{m+1}$. Calculating the first prolongation of
formulas (\ref{2.6}) we get the transformation rule for the first
derivatives of ${\mathbf v}$
\begin{equation}
\label{2.7}
{\mathbf v}'_x = {\mathbf W}(t,x,{\mathbf v},{\mathbf v}_x,\theta).
\end{equation}

Now we differentiate (\ref{1.6}) with respect to $x$ and make the
following change of dependent variables equations,
\begin{equation}
\label{2.8}
{\mathbf w} = {\mathbf v}_x,
\end{equation}
which yield
\begin{equation}
\label{2.9}
{\mathbf w}_{t}=\frac{\p}{\p x}\Bigl(\tilde{\mathbf f}(t,x,{\mathbf
w},\ldots,{\mathbf w}_{n-1})\Bigr).
\end{equation}
Formulas (\ref{2.6}), (\ref{2.7}) provide the image of the
transformation group (\ref{2.6}) under the mapping (\ref{2.7}), so
that
\begin{equation}
\label{2.10}
t' = T(t,x,{\mathbf v},\theta),\quad x' = X(t,x,{\mathbf
v},\theta),\quad {\mathbf w}'_x = {\mathbf W}(t,x,{\mathbf
v},{\mathbf w},\theta).
\end{equation}
Here ${\mathbf v} = \p_x^{-1}{\mathbf w}$.

Consequently, if one of the derivatives, ${\p T}/{\p v^i}, {\p
X}/{\p v^i}, {\p \mathbf W}/{\p v^i}$, does not vanish identically,
then (\ref{2.10}) is the nonlocal symmetry group of system of
evolution equations (\ref{2.9}).

The same line of reasonings applies to the case when system
(\ref{0.1}) admits partial CLR.

We summarize the above speculations in the form of the multi-step
algorithm for group classification of nonlocal symmetries of systems
of evolutions equations associated with a given system of the form
(\ref{0.1}).

Let system of evolution equations (\ref{0.1}) be invariant under
$N$-dimensional Lie symmetry algebra $\mathcal{L}_N$. For simplicity
we formulate the algorithm for the case of complete CLR.

\vspace{3mm}
\noindent
{\bf Algorithm 1.}\ {\em Classification of potential symmetries
of (\ref{0.1})}
\begin{enumerate}
\item{Calculate inequivalent subalgebras ${\mathcal{M}}$ of the
algebra $\mathcal{L}_N$.}
\item{Select those subalgebras ${\mathcal{M}}$, which contain
commutative subalgebras ${\mathcal{M}_m}$ of operators of the form
(\ref{1.3}).}
\item{For each commutative subalgebra ${\mathcal{M}_m}$ perform
change of variables reducing its basis elements to the canonical
forms $\p_{v^1}$, $\ldots$, $\p_{v^m}$ and transform correspondingly
initial system (\ref{0.1}).}
\item{Perform nonlocal transformation (\ref{2.8}).}
\item{Eliminate 'old' dependent variables ${\mathbf v}$ from
(\ref{2.6}) in order to derive symmetry group (\ref{2.10}) of the
transformed system of evolution equations (\ref{2.9}).}
\item{Verify that there is, at least, one derivative from the list
${\p T}/{\p v^i}$, ${\p X}/{\p v^i}$, ${\p \mathbf W}/{\p v^i}$ that
does not vanish identically. If this is the case, then (\ref{2.10})
is the nonlocal (potential) symmetry of (\ref{2.9}).}
\end{enumerate}

The steps needed to implement the above algorithm for the case of
system of evolution equations admitting partial CLR are the same,
the only difference is that intermediate formulas
(\ref{2.6})-(\ref{2.10}) are more cumbersome, since we need to
distinguish between two sets of dependent variables ${\mathbf u}$
and ${\mathbf w}$ (see, (\ref{1.2})).

Note that by force of Theorems 1,2 any potential symmetry of
equations of the form (\ref{0.1}) can be obtained in the above
described manner.

\section{Some generalizations}

Denote the class of partial differential equations of the form
(\ref{0.1}) as ${\mathfrak E}_n$. Then any system of the form
\begin{equation}
\label{3.1}
{\mathbf u}_t = f(t,x,{\mathbf u}_1,\ldots{\mathbf u}_n),
\end{equation}
$(i)$ belongs to ${\mathfrak E}_n$, and, $(ii)$ its image under
nonlocal transformation ${\mathbf u} = {\mathbf v}_x$ also belongs
to ${\mathfrak E}_n$. Existence of such nonlocal transformation is
in the core of our approach to classifying nonlocal symmetries of
systems of evolution equations.

It is not but natural to ask whether there are other types of
nonlocal transformations of the class ${\mathfrak E}_n$ that can be
utilized to generate nonlocal symmetries. Remarkably, such nonlocal
transformations do exist. Sokolov \cite{sok1} put forward the idea
of group approach to generating such transformations for a single
evolution equation. It is straightforward to modify his approach to
handle systems of evolution equations, as well. As an illustration,
we consider system (\ref{3.1}). It is invariant under the
$m$-dimensional Lie algebra $\mathcal{L}_m=\langle \p_{u^1},$
$\ldots,$ $\p_{u^m}\rangle$. The simplest set of $(m+2)$
functionally-independent invariants of the algebra $\mathcal{L}_m$
can be chosen as follows $t$, $x$, $u^1_x$, $\ldots$, $u^m_x$. Now
we define the transformation
\begin{eqnarray}
&&\bar t = T(t,x,{\mathbf u},{\mathbf u}_x,{\mathbf
u}_{xx},\ldots),\quad \bar x = X(t,x,{\mathbf u},{\mathbf
u}_x,{\mathbf u}_{xx},\ldots),\nn\\
&&\bar{\mathbf u} = {\mathbf U}(t,x,{\mathbf u},{\mathbf
u}_x,{\mathbf u}_{xx},\ldots)\label{3.2}
\end{eqnarray}
so that $T, X, {\mathbf U}$ are invariants of the symmetry group of
the system under study. In the case under consideration, we have
$T=t$, $X=x$, ${\mathbf U}={\mathbf u}_x$. As we established in
Section 1, applying this transformation to any equation of the form
(\ref{3.1}) yields system of evolution equations that belongs to
${\mathfrak E}_n$. What is more, Lie symmetry group of (\ref{3.1})
is mapped into symmetry group of the transformed system and some of
the basis operators of the latter become nonlocal ones.

Consider, as the next example system of evolution equations
\begin{equation}
\label{3.3}
{\mathbf u}_t = f(t,x,{\mathbf u}_2,\ldots{\mathbf u}_n),\quad n \ge
3.
\end{equation}
This system is invariant under the $2m$-dimensional Lie algebra
$\mathcal{L}_{2m}=\langle \p_{u^1}$, $\ldots$, $\p_{u^m}$,
$x\p_{u^1},$ $\ldots$, $x\p_{u^m}\rangle$. The simplest set of $m+2$
functionally independent first integrals reads as $t,x$, $u^1_{xx}$,
$\ldots$, $u^m_{xx}$. Consequently, change of variables (\ref{3.2})
takes the form
\begin{equation}
\label{3.4}
t=t,\quad x=x,\quad {\mathbf v}={\mathbf u}_{xx}.
\end{equation}
Note that we dropped the bars and replaced $\bar{\mathbf u}$ with
${\mathbf v}$.

Transforming (\ref{3.3}) according to (\ref{3.4}) we get
$$
\Bigl((\p_x^{-1})^2{\mathbf v}\Bigr)_t = f(t,x,{\mathbf v},{\mathbf
v}_1,\ldots{\mathbf v}_{n-2})
$$
or, equivalently,
$$
(\p_x^{-1})^2\Bigl({\mathbf v}_{t} - \p_x^2f(t,x,{\mathbf
v},{\mathbf v}_1,\ldots{\mathbf v}_{n-2})\Bigr) = 0.
$$
Integrating twice yields
\begin{equation}
\label{3.5}
{\mathbf v}_{t} = \p_x^2f(t,x,{\mathbf v},{\mathbf
v}_1,\ldots{\mathbf v}_{n-2}).
\end{equation}
Note that integration constants ${\mathbf w}^1(t)x+{\mathbf w}^2(t)$
are absorbed by the function ${\mathbf v}$.

So that nonlocal transformation (\ref{3.4}) maps a subset of
equations from ${\mathfrak E}_n$ into ${\mathfrak E}_n$.
Consequently, it can be used to generate nonlocal symmetries of the
initial system (\ref{3.3}).

Let system (\ref{3.3}) be invariant under the Lie transformation
group
\begin{equation}
\label{3.6}
t'=T(t,x,{\mathbf u},\theta),\quad x=X(t,x,{\mathbf u},\theta),\quad
{\mathbf u}={\mathbf U}(t,x,{\mathbf u},\theta).
\end{equation}
Computing the second prolongation of the above formulas we get the
transformation law for the functions ${\mathbf v}={\mathbf u}_{xx}$,
\begin{equation}
\label{3.7}
{\mathbf v}'={\mathbf V}(t,x,{\mathbf u}, {\mathbf u}_x, {\mathbf
v},\theta).
\end{equation}
Combining (\ref{3.6}) and (\ref{3.7}) yields the symmetry group of
system of evolution equations (\ref{3.5}),
\begin{equation}
\label{3.8}
t'=T(t,x,{\mathbf u},\theta),\quad x=X(t,x,{\mathbf u},\theta),\quad
{\mathbf v}'={\mathbf V}(t,x,{\mathbf u}, {\mathbf u}_x, {\mathbf
v},\theta)
\end{equation}
where ${\mathbf u}=(\p_x^{-1})^2{\mathbf v}$ are nonlocal variables.
Now, if one of the derivatives
$$
\frac{\p T}{\p u^i},\quad \frac{\p X}{\p u^i},\quad \frac{\p
{\mathbf V}}{\p u^i},\quad \frac{\p {\mathbf V}}{\p u^i_x}
$$
does not vanish identically, then (\ref{3.8}) is the nonlocal
symmetry group of system of evolution equations (\ref{3.5}).

It is important to emphasize that the symmetry algebra
$\mathcal{L}_m$ is not obliged to be commuting. The necessary
condition is that the corresponding transformation group has to
preserve the temporal variable, $t$, i.e., basis elements of
$\mathcal{L}_m$ have to be of the form
\begin{equation}
\label{3.9}
Q=\xi(t,x,{\mathbf u})\p_x + \sum_{j=1}^m\,\eta^j(t,x,{\mathbf
u})\p_{u^j}.
\end{equation}

As an illustration, consider the following system of second-order
evolution equations:
\begin{equation}
\label{3.10}
u^i_t=u^i_xf^i\left(t,x,\frac{u^1_{xx}}{u^1_x}, \ldots,
\frac{u^m_{xx}}{u^m_x}\right),\quad i=1,\ldots,m.
\end{equation}
This system is invariant under the $2m$-dimensional Lie algebra
$\mathcal{L}_{2m}=\langle\p_{u^1}$, $\ldots$, $\p_{u^m}$,
$u^1\p_{u^1}$, $\ldots$, $u^m\p_{u^m}\rangle$. Note that the algebra
$\mathcal{L}_{2m}$ is not commutative. The set of $m+2$ invariants
of the algebra $\mathcal{L}_{2m}$ can be chosen as follows,
$$
t,x,\frac{u^1_{xx}}{u^1_x}, \ldots, \frac{u^m_{xx}}{u^m_x}.
$$
making the change of variables
$$
t=t,\quad x=x,\quad v^1=\frac{u^1_{xx}}{u^1_x}, \ldots,
v^m=\frac{u^m_{xx}}{u^m_x}
$$
we rewrite (\ref{3.10}) in the form
\begin{equation}
\label{3.11}
\frac{\p}{\p t}\Bigl(\p_x^{-1}\exp(\p_x^{-1}v^i)\Bigr)
= \exp(\p_x^{-1}v^i)f^i(t,x,v^1,\ldots,v^m),\quad i=1,\ldots,m.
\end{equation}
Taking into account that the operators $\frac{\p}{\p t}$ and
$\p_x^{-1}$ commute, differentiating (\ref{3.11}) with respect to
$x$, and replacing ${\mathbf v}$ with ${\mathbf w}_x$ we finally get
\begin{equation}
\label{3.12}
w^i_t = w^i_xf^i(t,x,w^1_x, \ldots, w^m_x) + \frac{\p}{\p x}
f^i(t,x,w^1_x, \ldots, w^m_x),\quad i=1,\ldots,m.
\end{equation}
The above system is obtained from the initial one through the change
of dependent variables $u^i=\p_x^{-1}\exp(w^i),$\ $i=1,\ldots,m$.
Consequently, if system (\ref{3.10}) admits symmetry (\ref{3.6}),
then system (\ref{3.12}) admits the following transformation group:
\begin{equation}
\label{3.13}
t'=T(t,x,{\mathbf u},\theta),\quad x=X(t,x,{\mathbf u},\theta),\quad
{\mathbf w}={\mathbf W}(t,x,{\mathbf u}, {\mathbf w},\theta)
\end{equation}
with $u^i=\p_x^{-1}\exp(w^i),\ i=1,\ldots,m$. Again, if one of the
derivatives,
$$
\frac{\p T}{\p u^i},\quad \frac{\p X}{\p u^i},\quad \frac{\p
{\mathbf V}}{\p u^i},\quad \frac{\p {\mathbf W}}{\p u^i_x},
$$
does not vanish identically, then (\ref{3.13}) is the nonlocal
invariance group of system of evolution equations (\ref{3.11}).

The algorithm for group classification of nonlocal symmetries of
system (\ref{0.1}) suggested in the previous section yields those
nonlocal symmetries which are potential, since the nonlocal
transformation was chosen {\em a priori}. Allowing for a nonlocal
transformation to be determined by symmetry group of the system
under study, yields a more general algorithm for constructing
nonlocal symmetries.

Let system of evolution equations (\ref{0.1}) be invariant under
$N$-dimensional Lie symmetry algebra $\mathcal{L}_N$. Then the
following multi-step algorithm can used to construct nonlocal
symmetries of (\ref{0.1}).

\vspace{3mm}
\noindent
{\bf Algorithm 2.} {\em Classification of nonlocal symmetries
of (\ref{0.1})}
\begin{enumerate}
\item{Calculate inequivalent subalgebras ${\mathcal{M}}$ of the
algebra $\mathcal{L}_N$.}
\item{Select those subalgebras $\widetilde{\mathcal{M}}$, which
contain basis elements $e_1$, $\ldots$, $e_r$ of the form
(\ref{3.9}).}
\item{For each $\widetilde{\mathcal{M}}$ construct
$r+2$ functionally independent-invariants $\omega^t(t,x$, ${\mathbf
u},{\mathbf u}_x,\ldots)$, $\omega^x(t,x,{\mathbf u},{\mathbf
u}_x,\ldots)$, $\omega^1(t,x,{\mathbf u},{\mathbf u}_x,\ldots)$,
$\ldots$, $\omega^r(t,x,{\mathbf u},{\mathbf u}_x,\ldots)$ and make
change of variables
\begin{equation}
\label{3.14}
\bar t=\omega^t,\quad \bar x=\omega^x,\quad \bar u^i=\omega^i,\quad
i=1,\ldots,r.
\end{equation}
}
\item{Eliminate 'old' dependent variables ${\mathbf u}$ from
(\ref{3.14}) in order to derive symmetry group ${\mathcal G}$ of the
transformed system of evolution equations.} \item{Verify that there
is, at least, one function from the list $\{\omega^t$, $\omega^x$,
$\omega^1$, $\ldots$, $\omega^r\}$ that depends on $u^i$ for some
$1\le i\le r$. If this is the case, then ${\mathcal G}$ is the
nonlocal symmetry of (\ref{2.9}).}
\end{enumerate}

\section{Conclusion.}

One of the principal results of the paper is Theorem 3 from Section 2
stating that any potential symmetry of system of evolution equations
(\ref{0.1}) reduces to a Lie symmetry by an appropriate nonlocal
transformation of dependent and independent variables. The nonlocal
transformation in question is a superposition of the local change
of variables
\begin{equation}
\label{4.1}
\bar t=t,\quad \bar x=X(t,x,{\mathbf u}),\quad \bar{\mathbf
u}={\mathbf U}(t,x,{\mathbf u})
\end{equation}
and of the nonlocal change of dependent variables
\begin{equation}
\label{4.2}
{\mathbf v}=\bar{\mathbf u}_x.
\end{equation}
The explicit form of transformations (\ref{4.1}) is defined
by the Lie symmetry admitted by the corresponding system (\ref{0.1}).

we obtain as an by-product exhaustive characterization of systems
(\ref{0.1}), that can be represented in the form of conservation
law(s), in terms of Lie symmetries preserving the temporal variable,
$t$,
$$
t'=t,\quad x'=X(t,x,{\mathbf u},\theta),\quad {\mathbf u}'={\mathbf
U}(t,x,{\mathbf u},\theta)
$$
(see, Theorems 1,2).

In Section 3, we generalize the above reasonings in order to obtain
nonlocal symmetries which are not potential. The basic idea is
replacing (\ref{4.2}) with a more general nonlocal transformation.
This transformations is determined by invariants of Lie symmetry
algebra of the system under study.

We intend to devote one of our future publications to systematic study
of nonlocal symmetries of systems of nonlinear evolution equations
(\ref{0.1}) within the framework of the approach developed in Section 3.

\end{document}